\documentclass[nofootinbib,aps,prx,twocolumn,superscriptaddress,showpacs]{revtex4-1}
\usepackage{amsmath,amssymb,graphics,epsfig,epstopdf,color,verbatim,ulem,braket,tabularx}
\usepackage{multirow}
\usepackage[colorlinks,linkcolor=blue,citecolor=blue,urlcolor=blue,bookmarks=false]{hyperref}

\usepackage{listings}
\usepackage{cancel}
\usepackage{mathrsfs}
\usepackage{soul}
\usepackage{color}
\usepackage{url}
\usepackage{hyperref}

\newcommand{\beginsupplement}{%
        \setcounter{equation}{0}
        \renewcommand{\theequation}{S\arabic{equation}}%
        \setcounter{figure}{0}
        \renewcommand{\thefigure}{S\arabic{figure}}%
}

\makeatletter
\def\maketitle{
\@author@finish
\title@column\titleblock@produce
\suppressfloats[t]}
\makeatother

\begin{document}

\title{Maximum entropy analytic continuation of anomalous self-energies}

\author{Changming Yue}
\email{changming.yue@unifr.ch}
\affiliation{Department of Physics, University of Fribourg, 1700 Fribourg, Switzerland}

\author{Philipp Werner}
\email{philipp.werner@unifr.ch}
\affiliation{Department of Physics, University of Fribourg, 1700 Fribourg, Switzerland}

\begin{abstract}
The anomalous self-energy plays an important role in the analysis of superconducting states. Its spectral weight provides information on the pairing glue of superconductors, but it can change in sign. In many numerical approaches, for example Monte Carlo methods based on the Nambu formalism, the anomalous self-energy is obtained on the Matsubara axis, and nonpositive spectral weight cannot be directly obtained using the standard maximum entropy analytic continuation method. Here, we introduce an auxiliary self-energy corresponding to a linear combination of the normal and anomalous self-energies. We analytically and numerically prove that this auxiliary function has non-negative spectral weight independent of the pairing symmetry, which allows to compute the sign-changing spectrum of the original self-energy using the maximum entropy approach. As an application, we calculate the momentum-resolved spectral function of K$_3$C$_{60}$ in the superconducting state.
 \end{abstract}

\maketitle
 
\newpage

{\it Introduction.}
The anomalous self-energy $\Sigma^{\mathrm{ano}}(\bf{k},\omega)$, defined in the Gor'kov-Nambu formalism \cite{Nambu1960,Gorkov1958}, with $\bf{k}$ the crystal momentum and $\omega$ the real frequency, plays a crucial role in the theory of superconductivity. It  is proportional 
to the superconducting gap function 
$\Delta({\bf k},\omega)$ 
\cite{Gull2014},
which provides information on the spatial and dynamic structure of Cooper pairing. The spatial structure, encoded in the $\bf{k}$ dependence of $\Sigma^{\mathrm{ano}}$, determines the pairing symmetry as $s$, $p$, $d$ wave, etc. The static value $\Sigma^{\mathrm{ano}}{(\bf k},0)$ measures the strength of pairing, the high frequency limit $\Sigma^{\mathrm{ano}}{(\bf k},\infty)$ is proportional to the superconducting order parameter $\varDelta$, and the ratio $\Sigma^{\mathrm{ano}}({\bf k},0)/\varDelta$ defines the effective attractive interaction \cite{Yue2021}. The Cooper pairing in realistic superconductors is retarded. According to Migdal-Eliashberg theory \cite{Migdal1958,Eliashberg1960,Morel1961,McMillan1965,Carbotte1990,Chubukov2020,Marsiglio2020}, the retardation effect of conventional superconductors comes from the phonon dynamics, but shows up in the frequency-dependent electron self-energy. 
In unconventional superconductors, the retardation originates from other types of excitations, such as spin \cite{Scalapino1999,Hoshino2015}, orbital \cite{Kontani2010,Hoshino2017,Yue2021}, or nonlocal magnetic fluctuations \cite{Moriya2003,Poilblanc2005,Maier2008,Kyung2009}. 
The retardation effect is described by the frequency-dependent gap function, the calculation of which is closely related to that of $\Sigma^{\mathrm{ano}}(\bf{k},\omega)$. Obtaining the real-frequency $\Sigma^{\mathrm{ano}}$ is thus essential for understanding the properties of superconductors and in particular the pairing mechanisms. 

The retarded nature of the pairing can be experimentally measured. For example, the frequency-dependence of the local gap function can be measured in scanning tunneling microscopy (STM) experiments \cite{Giaever1962,Rowell1963,Rowell1971}. In the last two decades, the development of 
laser angle-resolved photoemission spectroscopy (laser ARPES) \cite{Damascelli2003,Liu2008} made it feasible to extract $\Sigma^{\mathrm{ano}}(\bf{k},\omega)$  and hence the gap function or pairing Eliashberg function 
 with high momentum and frequency resolution \cite{Shi2004,Bok2016}. Machine learning techniques have also recently been used to extract $\Sigma^{\mathrm{ano}}(\bf{k},\omega)$ from ARPES data \cite{Yamaji2021}. 

In numerical simulations of superconducting states, the normal and anomalous self-energies are usually obtained as a function of imaginary time or Matsubara frequency $i\omega_n$. Analytic continuation from the imaginary axis ($i\omega_n$) to the real frequency axis ($\omega+i\eta$) is needed to obtain the real-frequency self-energy. In Migdal-Eliashberg studies of conventional phonon-mediated paring, $\Sigma^{\mathrm{nor}}$ and $\Sigma^{\mathrm{ano}}$ can be obtained on the Matsubara axis with high numerical accuracy, so that the Pad\'e approximation can be used for the analytic continuation \cite{Vidberg1977}. Dynamical mean-field theory  \cite{MetznerVollhardt1989,GeorgesKotliar1992,Jarrell1992,Georges1996,Kotliar2006} and its cluster extensions \cite{Hettler1998,Hettler2000,Aryanpour2002,Maier2005,Lichtenstein2000,Kotliar2001,Liebsch2008} in the Gor'kov-Nambu formalism can be used to explore the typically unconventional superconducting states of strongly correlated electron systems. These methods map the interacting lattice system in the superconducting state to a self-consistently determined quantum impurity model with a superconducting (SC) bath. Such an impurity model with SC bath can be solved with the exact diagonalization (ED) method \cite{Caffarel1994,Kyung2009,Foley2019} directly on the real frequency axis, circumventing the problem of analytical continuation \cite{Sakai2016a,Sakai2016b,Sakai2018}. However,  ED impurity solvers can only treat a limited number of impurity and bath orbitals, which results in spiky spectra of the gap function or self-energy. 

Quantum Monte Carlo (QMC) methods provide a more accurate description of the unconventional pairing state on the Matsubara axis   \cite{Georges1993,Lichtenstein2000,Haule2007,Koga2010,Sentef2011,Sordi2012,Gull2013,Semon2014,Yue2022}. However, since QMC results contain noise in the Green's functions and self-energies, the Pad\'e approximation, which is not robust against noise, can become unreliable. This method also does not ensure the positive spectral weight of the normal self-energy. The maximum entropy (MaxEnt) method \cite{Jarrell1996} is a more robust and widely used method for the continuation of noisy numerical results, but its direct application is limited to functions with a positive definite spectrum. Unfortunately, the spectral weight of $\Sigma^{\mathrm{ano}}$ generally has sign changes on the real axis and thus MaxEnt cannot be directly applied. 

An auxiliary gap function $\tilde{\Delta}(i\omega_n)=[\Delta(i\omega_n)-\Delta(i0^+)]/(i\omega_n)$, which is odd in $\omega_n$,  was introduced in Ref.~\cite{Gull2014} and it was assumed that the spectral weight $-\mathrm{Im}\tilde{\Delta}(\omega+i\eta)$ is positive. However, ED results (\textit{e.~g.} Fig.~1 in Ref.~\onlinecite{Sakai2016a}) show that sign changes may appear in $-\mathrm{Im}{\Delta}(\omega+i\eta)$ and hence $-\mathrm{Im}\tilde{\Delta}(\omega+i\eta)$. In the case of particle-hole symmetric systems, one can introduce and analytically continue an auxiliary self-energy $\Sigma_\pm=\Sigma^{\mathrm{nor}}\pm\Sigma^{\mathrm{ano}}$ with non-negative spectral weight \cite{Gull2013b}, but this method does not apply to particle-hole asymmetric systems, such as cuprates with realistic band structures, or the fulleride compound K$_3$C$_{60}$ studied below.  Reymbaut {\it et al.} \cite{Reymbaut2015} proposed the so-called MaxEntAux method, where MaxEnt is applied to a properly defined auxiliary Green's function for a particle-hole mixed operator with positive-definite spectral weight $\mathcal{A}^{\operatorname{aux}}({\bf k}, \omega)\ge0$. In this method, the spectral function of the anomalous Green's function $\mathcal{A}^\text{ano}({\bf k}, \omega)$ can be  extracted as $\mathcal{A}^\text{ano}({\bf k}, \omega)=\frac{1}{2}\left[\mathcal{A}^{\operatorname{aux}}({\bf k}, \omega)-\mathcal{A}_{\uparrow}({\bf k}, \omega)-\mathcal{A}_{\downarrow}({\bf k},-\omega)\right]$, where $\mathcal{A}(\omega)$ is the normal spectral function. By using the Dyson equation, one can obtain from this the spectral function of the normal and anomalous self-energy, {\it if} the normal inter-orbital components of the lattice Green's function vanish. However, many realistic systems 
do not satisfy this condition  
for generic $\bf{k}$.
Also, to get the $\bf{k}$-resolved spectral function or optical conductivity in the SC state, one has to perform MaxEntAux for each ${\bf k}$ point, which is more time consuming than performing the analytic continuation of the self-energy, especially when the self-energy can be assumed to be local or restricted to a small cluster. 

{\it Auxiliary self-energy.}
The self-energy of an interacting system, regardless of whether it is local or non-local, can be 
reproduced by 
an auxiliary non-interacting Hamiltonian \cite{Balzer2014, Balzer2016,Seki2016}. This is achieved by connecting noninteracting bath sites to the sites of the lattice. The frequency dependence or retardation of the self-energy is mimicked by the hopping between the bath and lattice sites. (Note that this auxiliary bath is different from the effective bath of the DMFT impurity model, which represents the lattice environment.) Such an auxiliary model Hamiltonian is also used in the hidden fermion theory \cite{Sakai2016b}, where the $c$ electrons are the electrons of the system and the $f$ electrons (hidden fermions) are those of the auxiliary bath sites. The self-energy of the interacting system can be obtained by integrating out the $f$ degrees of freedom. In the SC state of the interacting system, the hidden fermion model reads \cite{Sakai2016b} 
\begin{align}
H_{cf}=&\sum_{\mathbf{k}\sigma}\Big\{ \left[\epsilon_{c}(\mathbf{k})+s(\mathbf{k})\right]c_{\mathbf{k}\sigma}^{\dagger}c_{\mathbf{k}\sigma}+\sum_{\alpha}\epsilon_{f_{\alpha}}(\mathbf{k})f_{\alpha\mathbf{k}\sigma}^{\dagger}f_{\alpha\mathbf{k}\sigma}\Big\} \nonumber\\
&+\sum_{\mathbf{k}\sigma\alpha}V_{\alpha}(\mathbf{k})\left(f_{\alpha\mathbf{k}\sigma}^{\dagger}c_{\mathbf{k}\sigma}+c_{\mathbf{k}\sigma}^{\dagger}f_{\alpha\mathbf{k}\sigma}\right)\nonumber\\
&-\sum_{\mathbf{k}}D_{c}(\mathbf{k})\left(c_{\mathbf{k}\uparrow}c_{-\mathbf{k}\downarrow}+\text{H.c.}\right)\nonumber\\
&-\sum_{\alpha\mathbf{k}}D_{f_{\alpha}}(\mathbf{k})\left(f_{\alpha\mathbf{k}\uparrow}f_{\alpha-\mathbf{k}\downarrow}+\text{H.c.}\right),
\label{eq:Hcf}
\end{align}
where $\epsilon_{c}(\mathbf{k})$ and $\epsilon_{f_\alpha}(\mathbf{k})$ are the bare dispersions of the $c$ and $f$ fermions, while $D_{c}(\mathbf{k})$ and $D_{f_\alpha}(\mathbf{k})$ represent the pairing strength between the $c$ and $f$ fermions, respectively.
The pairing symmetry is encoded in the $\mathbf{k}$-dependence of $D_{c}(\mathbf{k})$ and $D_{f_\alpha}(\mathbf{k})$. 
In Eq.~(\ref{eq:Hcf}), we consider the one-band case for the $c$ fermions. 

Let us introduce the notations $\omega^+\equiv \omega+i\eta$ [$(-\omega)^+\equiv -\omega+i\eta$ ] with $\eta=0^+$. By integrating out the $f$ degrees of freedom, one obtains the normal and anomalous self-energy of the $c$ fermion as 
\begin{align}
&\Sigma_{\bf k}^{\mathrm{nor}}(\omega^+)=s(\mathbf{k})+\sum_{\alpha}\frac{V_{\alpha}(\mathbf{k})^{2}\left[\omega^+ +\epsilon_{f_{\alpha}}(\mathbf{k})\right]}{(\omega^+)^{2}-\epsilon_{f_{\alpha}}(\mathbf{k})^{2}-D_{f_{\alpha}}(\mathbf{k})^{2}},\label{eq:Sigma_cf_nor}\\
&\Sigma_{\bf k}^{\mathrm{ano}}(\omega^+)=D_{c}(\mathbf{k})+\sum_{\alpha}\frac{-V_{\alpha}(\mathbf{k})^{2}D_{f_{\alpha}}(\mathbf{k})}{(\omega^+)^{2}-\epsilon_{f_{\alpha}}(\mathbf{k})^{2}-D_{f_{\alpha}}(\mathbf{k})^{2}}.
\label{eq:Sigma_cf}
\end{align}
The derivation of these expressions is provided in Appendix A of Ref.~\onlinecite{Sakai2016b}. $s(\mathbf{k})$ [$D_{c}(\mathbf{k}$)] is the frequency-independent part of the normal (anomalous) self-energy. Obviously, the poles of $\Sigma_{\bf k}^{\mathrm{nor}}$ and $\Sigma_{\bf k}^{\mathrm{ano}}$ are located at the same energies $\omega_{\bf k}=\pm \sqrt{\epsilon_{f_{\alpha}}(\mathbf{k})^2+D_{f_{\alpha}}(\mathbf{k})^2}$. 

We now define the auxiliary self-energy in the Matsubara frequency space as the following linear combination of the normal and anomalous self-energy,
\begin{align}
\Sigma^{\mathrm{aux}}_{\bf k}(i\omega_{n})&=\Sigma^{\mathrm{ano}}_{\bf k}(i\omega_{n})+\frac{\Sigma^{\mathrm{nor}}_{{\bf k},\uparrow}(i\omega_{n})-\Sigma^{\mathrm{nor}}_{{\bf k},\downarrow}(-i\omega_{n})}{2}\nonumber\\
&=\Sigma^{\mathrm{ano}}_{\bf k}(i\omega_{n})+i\mathrm{Im}\Sigma^{\mathrm{nor}}_{\bf k}(i\omega_{n}),
\label{eq:Saux_matsu}
\end{align}
where in the second step a paramagnetic state is assumed with $\Sigma^{\mathrm{nor}}_{{\bf k},\uparrow}=\Sigma^{\mathrm{nor}}_{{\bf k},\downarrow}=\Sigma^{\mathrm{nor}}_{{\bf k}}$. 
Up to a factor $\frac{1}{2}$, Eq.~(\ref{eq:Saux_matsu}) corresponds to the sum of all the elements of the self-energy matrix in the Nambu formalism, which is similar to the auxiliary Green's function $G^{\operatorname{aux}}_{\bf k}(i \omega_{n})=G^{\mathrm{nor}}_{{\bf k},\uparrow}(i \omega_{n})-G^{\mathrm{nor}}_{{\bf k},\downarrow}(-i \omega_{n})+2 G^{\mathrm{ano}}_{\bf k}( i \omega_{n})$ introduced by Reymbaut {\it et al.} in the MaxEntAux approach \cite{Reymbaut2015}.  

The real frequency auxiliary self-energy of the paramagnetic state thus reads
\begin{align}
&\Sigma^{\mathrm{aux}}_{\bf k}(\omega^+)=\Sigma^{\mathrm{ano}}_{\bf k}(\omega^+)+\frac{\Sigma^{\mathrm{nor}}_{\bf k}(\omega^+)-\Sigma^{\mathrm{nor}}_{\bf k}(-\omega^+)}{2} \nonumber\\
&=\mathrm{Re}\Sigma_{\bf k}^{\mathrm{ano}}(\omega^{+})+\frac{\mathrm{Re}\Sigma_{\bf k}^{\mathrm{nor}}(\omega^{+})-\mathrm{Re}\Sigma_{\bf k}^{\mathrm{nor}}[(-\omega)^{+}]}{2}\nonumber\\
&+i\mathrm{Im}\Sigma_{\bf k}^{\mathrm{ano}}(\omega^{+})+i\frac{\mathrm{Im}\Sigma_{\bf k}^{\mathrm{nor}}(\omega^{+})+\mathrm{Im}\Sigma_{\bf k}^{\mathrm{nor}}[(-\omega)^{+}]}{2}
\label{eq:Saux_realfr}
\end{align}
where the property $f(-\omega^+)=f(-\omega-i\eta)=f(-\omega+i\eta)^{*}=f[(-\omega)^+]^*$ has been used. 

The auxiliary self-energy corresponding to Eqs.~(\ref{eq:Sigma_cf_nor}) and (\ref{eq:Sigma_cf})  has positive definite spectral weight. 
A straightforward derivation [see Supplementary Material (SM)] gives 
\begin{align}
&\mathrm{Im}\Sigma^{\mathrm{aux}}(\omega+i\eta)=-\sum_{\alpha}V_{\alpha}(\mathbf{k})^{2}\eta\nonumber\\
&\hspace{5mm}\times\frac{(\omega-D_{f_{\alpha}}(\mathbf{k}))^{2}+\eta^{2}+\epsilon_{f_{\alpha}}(\mathbf{k})^{2}}{\left[\omega^{2}-\eta^{2}-\epsilon_{f_{\alpha}}(\mathbf{k})^{2}-D_{f_{\alpha}}(\mathbf{k})^{2}\right]^{2}+4\omega^{2}\eta^{2}}<0,
\end{align}
which yields $A_\Sigma^{\mathrm{aux}}(\omega)=-\mathrm{Im}\Sigma^{\mathrm{aux}}(\omega+i\eta)/\pi>0$. Hence, the auxiliary self-energy defined in Eq.~(\ref{eq:Saux_matsu}) has positive spectral weight, and thus can be analytically continued with MaxEnt.  

If the normalization of $A_\Sigma^{\mathrm{nor}}(\omega)$ and $A_\Sigma^{\mathrm{aux}}(\omega)$ can be determined by calculating static four-particle correlation functions \cite{Wang2011}, it is possible to perform MaxEnt directly on $\Sigma^{\mathrm{nor}}(i\omega_n)$ and $\Sigma^{\mathrm{aux}}(i\omega_n)$. Alternatively, one may define two auxiliary Green's functions for $\Sigma^{\mathrm{aux}}$ and $\Sigma^{\mathrm{nor}}$, 
\begin{align}
&\tilde{G}^{\mathrm{aux}}_{1,\bf k}(i\omega_n)=1/\left[i\omega_n-\left[\Sigma^{\mathrm{nor}}_{\bf k}(i\omega_n)-\Sigma^{\mathrm{nor}}_{\bf k,\infty}\right]\right],\nonumber\\
&\tilde{G}^{\mathrm{aux}}_{2,\bf k}(i\omega_n)=1/\left[i\omega_n-\left[\Sigma^{\mathrm{aux}}_{\bf k}(i\omega_n)-\Sigma^{\mathrm{aux}}_{\bf k,\infty}\right]\right],
\end{align}
respectively. Since both $\tilde{G}^{\mathrm{aux}}_{1,\bf k}(i\omega_n)$ and $\tilde{G}^{\mathrm{aux}}_{2,\bf k}(i\omega_n)$ have positive definite spectral weight, we can perform a standard MaxEnt analytic continuation \cite{Jarrell1996} and then apply the Kramers-Kronig transformation to get $\tilde{G}^{\mathrm{aux}}_{1,\bf k}(\omega+i\eta)$ and  $\tilde{G}^{\mathrm{aux}}_{2,\bf k}(\omega+i\eta)$. We note that MaxEnt is not the only choice for the continuation of $\tilde{G}^{\mathrm{aux}}_{2,\bf k}(i\omega_n)$. All analytic continuation methods taking into account the positivity of the spectral weight can be used, including MaxEnt \cite{Jarrell1996},  Sparse Modeling (SpM) \cite{Otsuki2017,Yoshimi2019}, an artificial neural network approach \cite{Fournier2020} for numerical results with stochastic noise, and the Nevanlinna analytical continuation method \cite{Fei2021}. Also the Pad\'e approximation \cite{Vidberg1977} can in principle be used if the stochastic noise is small enough. 
After this step, we can obtain the real-frequency self-energy as follows:
\begin{align}
\Sigma^{\mathrm{nor}}_{\bf k}(\omega^+)&=\omega^+ - \tilde{G}^{\mathrm{aux}}_{1,\bf k}(\omega^+)^{-1} + \Sigma^{\mathrm{nor}}_{\bf k,\infty},\nonumber\\
\Sigma^{\mathrm{aux}}_{\bf k}(\omega^+)&=\omega^+ - \tilde{G}^{\mathrm{aux}}_{2,\bf k}(\omega^+)^{-1} + \Sigma^{\mathrm{aux}}_{\bf k,\infty},\nonumber\\
\Sigma^{\mathrm{ano}}_{\bf k}(\omega^+)&=\Sigma^{\mathrm{aux}}_{\bf k}(\omega^+)-\frac{\Sigma^{\mathrm{nor}}_{\bf k}(\omega^+)-\Sigma^{\mathrm{nor}}_{\bf k}(-\omega^+)}{2}.
\end{align}

{\it Benchmarks.} 
In the following, we provide two benchmarks to verify that $A_\Sigma^{\mathrm{aux}}(\omega)$ is positive definite. 

(1) $s$-wave spin-singlet pairing. Here, we consider an Anderson impurity model with a single bath site with $s$-wave pairing. The Hamiltonian is
\begin{align}
H=&\, U\hat{n}_{d\uparrow}\hat{n}_{d\downarrow}-\mu(\hat{n}_{d\uparrow}+\hat{n}_{d\downarrow})+\epsilon_{b}c_{\uparrow}^{\dagger}c_{\uparrow}+\epsilon_{b}c_{\downarrow}^{\dagger}c_{\downarrow}\nonumber\\
&+\Delta(c_{\uparrow}c_{\downarrow}+h.c.)+(Vd_{\uparrow}^{\dagger}c_{\uparrow}+Vd_{\downarrow}^{\dagger}c_{\downarrow}+h.c.),
\label{eq:H_1band1bath}
\end{align}
with $d$ the impurity operators and $c$ the bath operators. $U$ is the onsite interaction, $\mu$ the chemical potential, $V$ the hybridization parameter, $\epsilon_b$ the bath energy level and $\Delta$ the pair field.  The real-frequency normal and anomalous Green's functions can be calculated using the Lehmann representation, with eigenstates and eigenenergies obtained by ED. The Dyson equation in the Nambu formalism then yields the self-energy. The real-frequency auxiliary self-energy is obtained according to Eq.~(\ref{eq:Saux_realfr}). As shown in Fig.~\ref{fig:Fig1}, which corresponds to a system away from half-filling, $\mathrm{Im}\Sigma^{\mathrm{aux}}(\omega+i\eta)$ is negative for all real frequencies. 

\begin{figure}[t]
\includegraphics[clip,width=2.6in,angle=0]{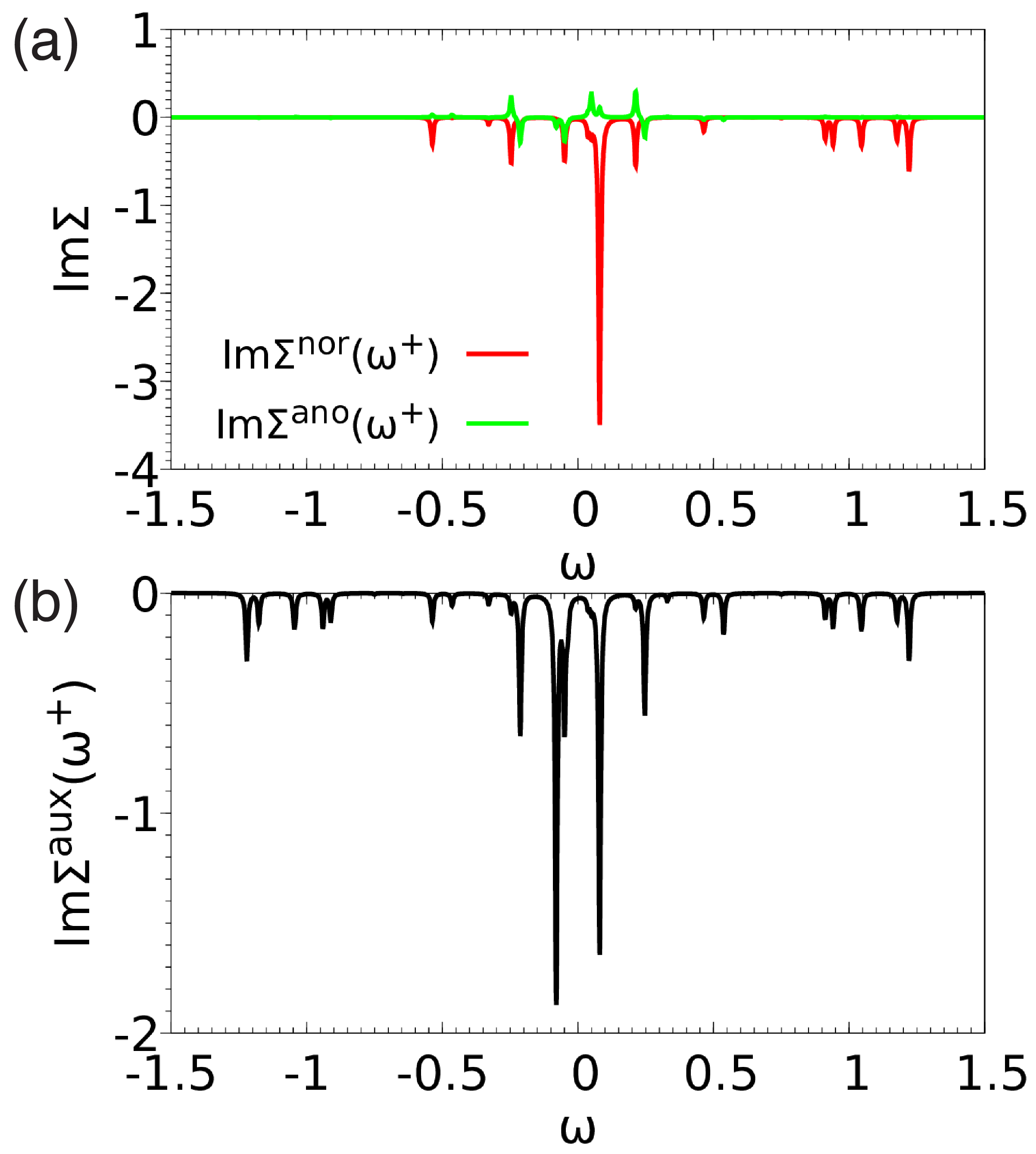}
\caption{
(a) [(b)] Imaginary part of the real-frequency normal and anomalous (auxiliary) self-energy corresponding to an impurity system with $s$-wave pairing (Eq.~(\ref{eq:H_1band1bath})). The parameters are $\epsilon_b=0$, 
$V=0.2$, $T=0.1$, $U=1.0$, $\Delta=0.05$, $\mu=0.5U-0.4$, $\eta=0.005$. The total filling of the system (impurity plus bath site) is 1.77, while half-filling corresponds to 2. 
}
\label{fig:Fig1}
\end{figure}

(2) $d$-wave spin singlet pairing. In the $2\times2$ cellular DMFT description with ED impurity solver, the $d$-wave SC state of the single band Hubbard model on the square lattice is mapped to a four-site cluster impurity model with, \textit{e.~g.}, two bath orbitals coupled to each site \cite{Kancharla2008,Civelli2009,Kyung2009,Foley2019,Sakai2016a}. The cluster Green's function in the Nambu formalism is diagonal in the four cluster momenta $\Gamma=(0,0)$, $X=(\pi,0)$, $Y=(0,\pi)$ and $M=(\pi,\pi)$. In the SC state, only the anomalous self-energies for $X$ and $Y$ are non-zero, and they are opposite in sign. We choose the $X$ point and show the results for filling $n=0.935$ per site in Fig.~\ref{fig:Fig2}. As can be seen in panel (b), the spectral weight of the auxiliary self-energy $-\text{Im}\Sigma^{\mathrm{aux}}(\omega^+)/\pi$ is positive. 
\begin{figure}
\includegraphics[clip,width=2.6in,angle=0]{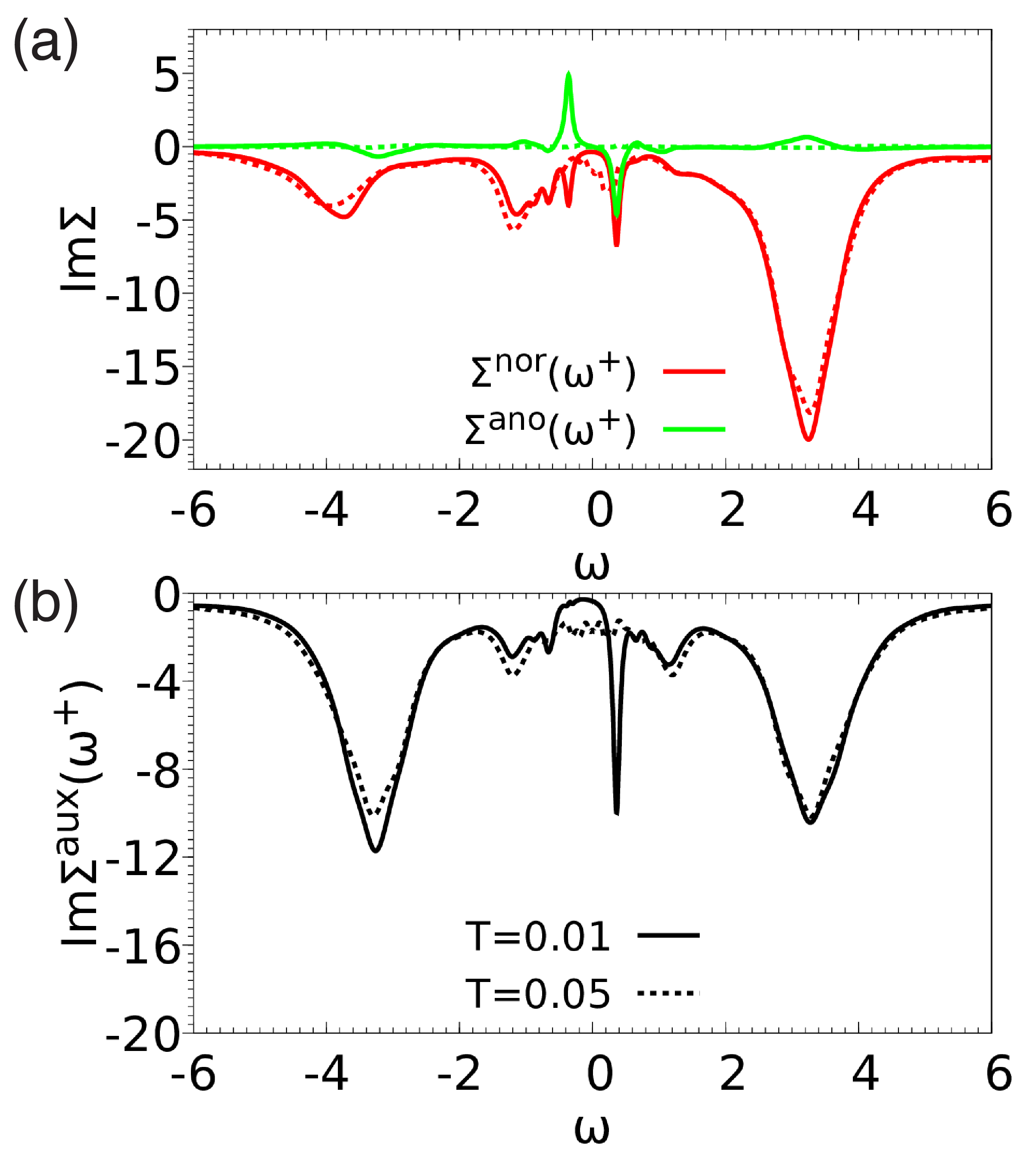}
\caption{
(a)[(b)] The imaginary part of the real-frequency normal and anomalous (auxiliary) self-energy at the $X$ point for a $d$-wave SC solution of the Hubbard model, calculated within CDMFT with an ED impurity solver.  The filling per site is 0.935 and the interaction is $U=8$. The data of the normal and anomalous self-energy are reproduced from Fig.~4 of Ref.~\cite{Sakai2016b}. The solid (dashed) line shows the result deep blow (close to) $T_c$. 
 }
\label{fig:Fig2}
\end{figure}

{\it Application to the superconducting state of K$_3$C$_{60}$.}  The alkali-doped fullerides A$_3$C$_{60}$ (A=K, Rb, Cs) are strongly correlated electron systems which exhibit a SC dome next to a Mott insulating phase \cite{Hebard1991,Crespi2002,Capone2002,Capone2009,Nomura2012,Nomura2015,Zadik2015,Nomura2016,Prassides2016}. They are described by a three-band Hubbard model with an inverted Hund's coupling resulting from the overscreening of the small bare Hund's coupling by Jahn-Teller phonons \cite{Fabrizio1997,Capone2002,Nomura2015}. The intra-orbital spin-singlet SC state is unconventional, with a pairing glue originating from local orbital fluctuations, according to DMFT studies \cite{Steiner2016,Hoshino2017,Yue2021}.

To the best of our knowledge, there are no experimental ARPES results for the SC state of A$_3$C$_{60}$, possibly due to the strong air-sensitivity of these compounds and the lack of good single crystals. We are also not aware of any calculated momentum-resolved electronic structures for the SC state of A$_3$C$_{60}$. The analytic continuation method presented in this work makes it possible to obtain the real-frequency anomalous self-energy, gap function and $A({\bf k},\omega)$.

Here, we focus on K$_3$C$_{60}$ which is superconducting below $T_c \approx $ 18 K \cite{Hebard1991}. The interaction parameters are obtained using the constrained random phase approximation (cRPA) \cite{Aryasetiawan2004}, which yields $U_{\mathrm{cRPA}}= 0.8552$ eV and $J_{\mathrm{cRPA}}=0.0378$ eV, roughly consistent with the values in Ref.~\onlinecite{Nomura2012}. Considering the effect of phonon screening on the interactions within constrained density functional perturbation theory \cite{Nomura2015_cDFPT}, the effective interaction parameters become $U_{\mathrm{eff}}=0.703$ eV and $J_{\mathrm{eff}}=-0.0122$ eV. We solve the realistic three $t_{1u}$ band Hubbard model with rotationally invariant Kanamori interactions in the framework of density functional theory (DFT) plus DMFT \cite{Georges1996,Kotliar2006}. To deal with the $s$-wave intra-orbital spin-singlet pairing, we implement DFT+DMFT in the Nambu formalism \cite{Nomura2015} and solve the corresponding three-orbital Anderson impurity model with normal and anomalous hybridization functions using continuous-time quantum Monte Carlo simulations in the hybridization expansion (CT-HYB) \cite{Werner2006,Gull2011}. To ensure the ergodicity of the Monte Carlo sampling, we also implemented four-operator updates \cite{Semon2014,Yue2022} in CT-HYB.

\begin{figure}[htp]
\includegraphics[clip,width=3.4in,angle=0]{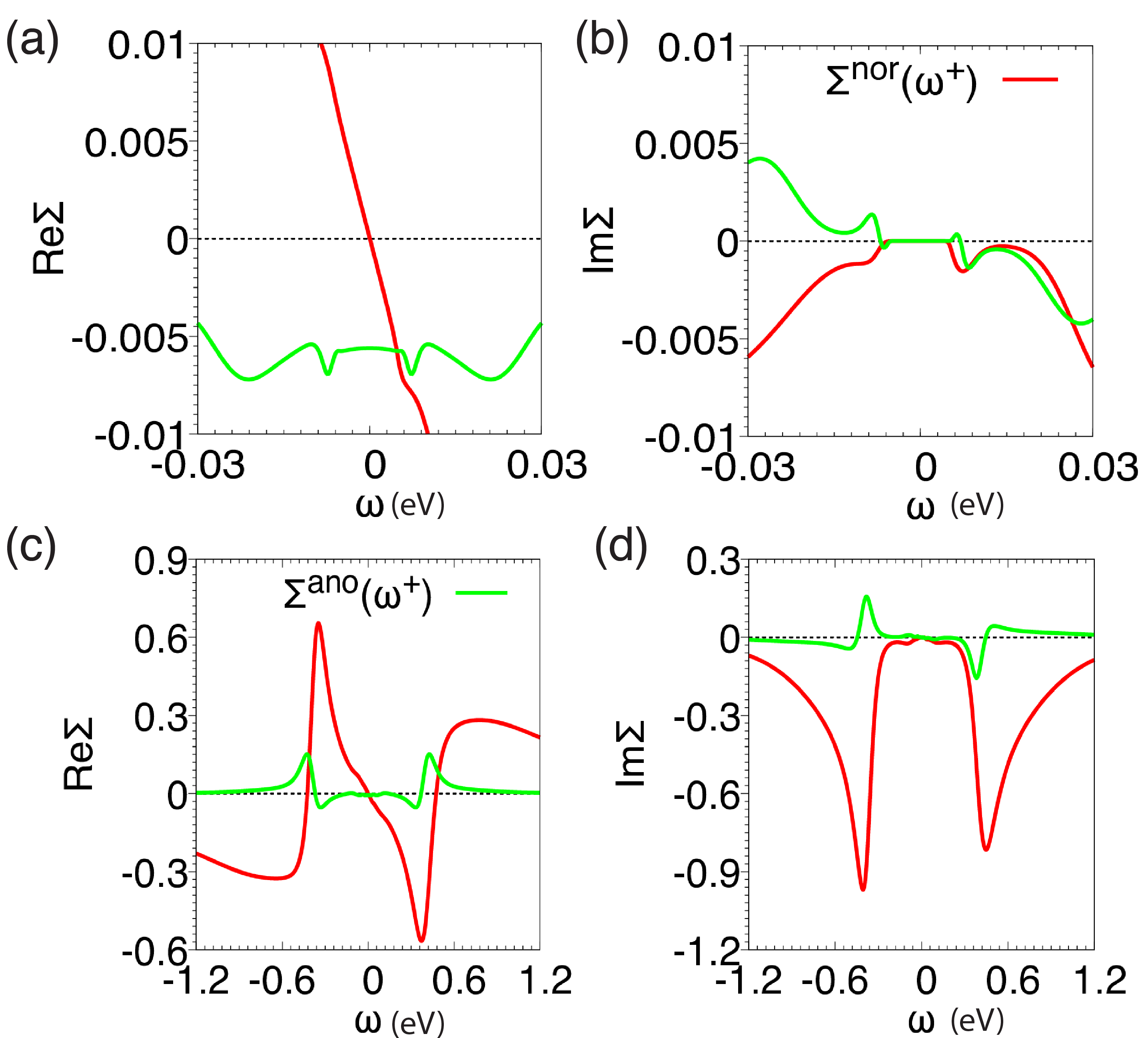} 
\caption{
The real frequency $\Sigma^{\mathrm{nor}}(\omega)$ and $\Sigma^{\mathrm{ano}}(\omega)$ of K$_3$C$_{60}$ at $T=10$~K obtained by the auxiliary analytic continuation method.
Panels (a,b) [(c,d)] show the results on a smaller (larger) energy window. $\mathrm{Re}\Sigma^{\mathrm{nor}}(\omega)$ (red line) in panel (a) is shifted down by the Hartree-Fock value $\mathrm{Re}\Sigma^{\mathrm{nor}}_{\infty}$=1.82 for a better visualization. The corresponding Matsubara frequency $\Sigma^{\mathrm{nor}}(i\omega_n)$ and $\Sigma^{\mathrm{ano}}(i\omega_n)$ are provided in the SM.
Here, we only show results for one of the degenerate $t_{1u}$ orbitals. 
 }
\label{fig:Fig3}
\end{figure}

The real-frequency $\Sigma^{\mathrm{nor}}(\omega)$ and $\Sigma^{\mathrm{ano}}(\omega)$ at $T=10$~K in the SC state, obtained with the auxiliary analytic continuation method and the MaxEnt code of Ref.~\onlinecite{Levy2017}, are plotted in Fig.~\ref{fig:Fig3}. As shown by the green curves, $\Sigma^{\mathrm{ano}}(\omega)$ features sign changes both in the real and imaginary parts. The corresponding $A({\bf k},\omega)$ is presented in Fig.~\ref{fig:Fig4}. Panel (a) shows the results for $T=10$~K ($<T_c\approx$ 18 K), and panel (b) a zoom of the low-energy spectrum along  the $\mathrm{\Gamma - X}$ path. We find a tiny SC gap with $\Delta_{\mathrm{sc}}\approx 7.31/2=3.66$~meV, extracted at the half peak-height \cite{Yue2021} (panel (d)), which is slightly larger than the experimental value of 3.0~meV extracted from optical conductivity measurements 
\cite{Degiorgi1994}. 
The ratio $2\Delta_{\mathrm{sc}}/k_BT_c \approx 4.71$ is 
 in good agreement with the value recently extracted from Raman scattering in Ref.~\cite{Wang2023}. 
A $2\Delta_{\mathrm{sc}}/k_BT_c$ ratio not much larger than the BCS result is consistent with the fact that K$_3$C$_{60}$ is on the weak-coupling side of the SC dome \cite{Zadik2015}. 

As one can see from Fig.~\ref{fig:Fig4}(b), faint spectral features exist inside the gap. This residual spectral weight leads to a partial filling-in of the gap in the local spectral function, as seen in panel (d). It also contributes to the optical conductivity, which is consistent with the experimentally observed upturn in the real part of the optical conductivity in the energy range $< 5$~meV, which grows as temperature is raised from  $T=6$ K to $T=T_c$ \cite{Degiorgi1994}. 
If $T$ is increased above $T_c$, the gap disappears, see the spectrum for 30~K in panel (c). Furthermore, the bands broaden, which shows that the system becomes less coherent. Indeed, according to previous model studies, we expect the system to approach an orbital freezing crossover at even higher temperatures \cite{Hoshino2017}. The ${\bf k}$-resolved spectra in a larger energy window, which also show the Hubbard bands, can be found in the SM.

\begin{figure*}[htp]
\includegraphics[clip,width=0.8\paperwidth,angle=0]{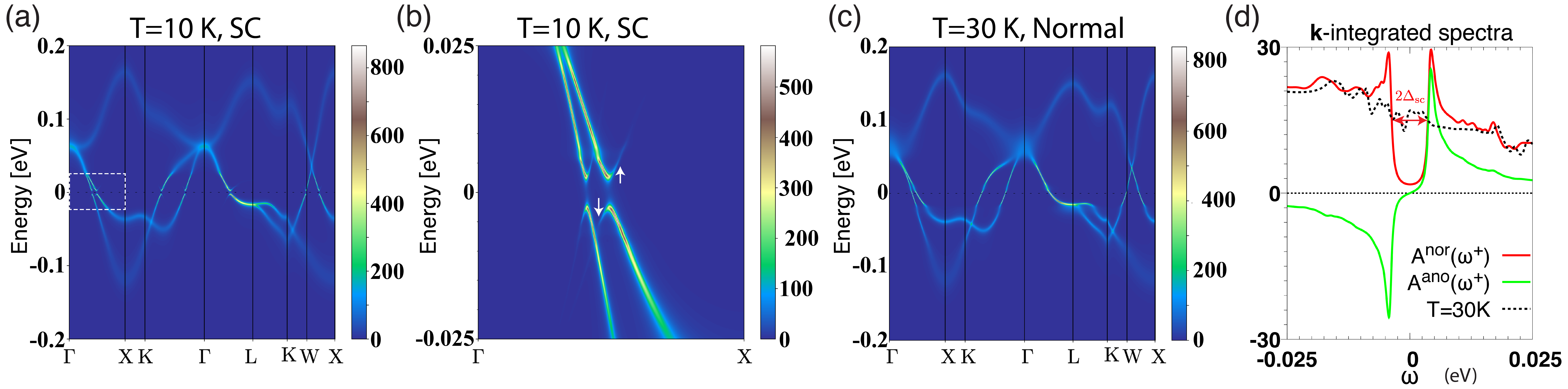}
\caption{
Momentum-resolved spectral function $A({\bf k},\omega)$ of K$_3$C$_{60}$. 
(a) $A({\bf k},\omega)$ for the superconducting state at $T=10$~K.
(b) Zoom of the low-energy spectrum along $\mathrm{\Gamma - X}$ in the region marked by the white dashed box in (a).  The white arrows highlight the back-bending of the Bogoliubov bands.
(c) $A({\bf k},\omega)$ for the normal state at $T=30$~K.
(d) Low energy structures of the ${\bf k}$-integrated normal and anomalous spectra $A^\text{nor(ano)}(\omega)=-\frac{1}{\pi}\text{Im}\frac{1}{N_{\bf k}}\sum_{\bf k}G^\text{nor(ano)}(\bf{k},\omega)$.
}
\label{fig:Fig4}
\end{figure*}

{\it Conclusions.}
We have addressed the problem of analytic continuation of the anomalous self-energy with non-positive spectral weight. We introduced an auxiliary self-energy, $\Sigma^{\mathrm{aux}}$, which is a linear combination of the normal and anomalous self-energy, and showed both analytically and numerically that this function 
has positive definite spectral weight, regardless of the space-time or spin structure of the pairing state. This allows us to use the MaxEnt method to analytically continue $\Sigma^{\mathrm{aux}}$ from the Matsubara to the real-frequency axis and to extract the spectral weight of $\Sigma^{\mathrm{ano}}$. Our method paves the way for systematic studies of the momentum-resolved electronic structures of conventional and unconventional superconductors. Access to the real-frequency anomalous self-energy makes it possible to calculate the gap function, spectral function  and optical conductivity, which are important for comparisons with experiments, and for revealing the pairing mechanism. We have demonstrated the usefulness of our approach by computing spectral function of K$_3$C$_{60}$ in the SC state. In the future, it will be interesting to extend these calculations to other types of superconductors, and in particular to the $d$-wave superconducting state of cuprates. 

{\it Acknowledgements. ---}
We thank S. Sakai for providing the ED data used in Fig.~\ref{fig:Fig3} and A.M.S Tremblay for helpful discussions. The calculations were performed on the Beo05 cluster at the University of Fribourg, using a code based on iQIST  \cite{HUANG2015140,iqist}. C.Y. and P.W. acknowledge support from SNSF Grant No.~200021-196966. The data presented in this work can be downloaded from \href{https://doi.org/10.5281/zenodo.7779381}{https://doi.org/10.5281/zenodo.7779381}.

\clearpage

\beginsupplement
\title{Supplementary Information for\\ Maximum entropy analytic continuation of anomalous self-energies}
\maketitle

\begin{widetext}
\begin{flushleft}

\section*{SM1. Positive Definiteness of the Spectral Weight of the Auxiliary Self-Energy}

In this part, we prove that $\Sigma^{\mathrm{aux}}$ has positive
spectral weight, i.e. $A_{\Sigma}^{\mathrm{aux}}(\omega)=-\mathrm{Im}\Sigma^{\mathrm{aux}}(\omega+i\eta)/\pi>0$.
For convenience, we define the real function $\xi(\omega,{\bf k})\equiv\omega^{2}-\eta^{2}-\epsilon_{f_{\alpha}}(\mathbf{k})^{2}-D_{f_{\alpha}}(\mathbf{k})^{2}=\xi(-\omega,{\bf k})$ and first rewrite the normal and anomalous self-energy using  $\xi(\omega,{\bf k})$.\\
\mbox{}\\
(1) The normal self-energy reads 
\begin{align}
\Sigma_{{\bf k}}^{\mathrm{nor}}(\omega+i\eta) & =s(\mathbf{k})+\sum_{\alpha}\frac{V_{\alpha}(\mathbf{k})^{2}\left[\omega+i\eta+\epsilon_{f_{\alpha}}(\mathbf{k})\right]}{\left[(\omega+i\eta)^{2}-\epsilon_{f_{\alpha}}(\mathbf{k})^{2}-D_{f_{\alpha}}(\mathbf{k})^{2}\right]}\nonumber\\
 & =s(\mathbf{k})+\sum_{\alpha}\frac{V_{\alpha}(\mathbf{k})^{2}\left[\omega+i\eta+\epsilon_{f_{\alpha}}(\mathbf{k})\right]}{\omega^{2}-\eta^{2}-\epsilon_{f_{\alpha}}(\mathbf{k})^{2}-D_{f_{\alpha}}(\mathbf{k})^{2}+i2\omega\eta}\nonumber\\
 & =s(\mathbf{k})+\sum_{\alpha}\frac{V_{\alpha}(\mathbf{k})^{2}\left[\omega+i\eta+\epsilon_{f_{\alpha}}(\mathbf{k})\right]}{\xi(\omega,{\bf k})+i2\omega\eta}\nonumber\\
 & =s(\mathbf{k})+\sum_{\alpha}\frac{V_{\alpha}(\mathbf{k})^{2}\left[\omega+i\eta+\epsilon_{f_{\alpha}}(\mathbf{k})\right]\left[\xi(\omega,{\bf k})-i2\omega\eta\right]}{\xi(\omega,{\bf k})^{2}+4\omega^{2}\eta^{2}}.
\label{eq:sm_Snor}
\end{align}
(2) The anomalous self-energy reads 
\begin{align}
\ensuremath{\Sigma_{{\bf k}}^{\mathrm{ano}}(\omega+i\eta)} & =D_{c}(\mathbf{k})+\sum_{\alpha}\frac{-V_{\alpha}(\mathbf{k})^{2}D_{f_{\alpha}}(\mathbf{k})}{(\omega+i\eta)^{2}-\epsilon_{f_{\alpha}}(\mathbf{k})^{2}-D_{f_{\alpha}}(\mathbf{k})^{2}}\nonumber\\
 & =D_{c}(\mathbf{k})+\sum_{\alpha}\frac{-V_{\alpha}(\mathbf{k})^{2}D_{f_{\alpha}}(\mathbf{k})}{\omega^{2}-\eta^{2}-\epsilon_{f_{\alpha}}(\mathbf{k})^{2}-D_{f_{\alpha}}(\mathbf{k})^{2}+i2\omega\eta}\nonumber\\
 & =D_{c}(\mathbf{k})+\sum_{\alpha}\frac{-V_{\alpha}(\mathbf{k})^{2}D_{f_{\alpha}}(\mathbf{k})}{\xi(\omega,{\bf k})+i2\omega\eta}\nonumber\\
 & =D_{c}(\mathbf{k})+\sum_{\alpha}\frac{-V_{\alpha}(\mathbf{k})^{2}D_{f_{\alpha}}(\mathbf{k})\left[\xi(\omega,{\bf k})-i2\omega\eta\right]}{\xi(\omega,{\bf k})^{2}+4\omega^{2}\eta^{2}}.
\label{eq:sm_Sano}
\end{align}

$\Sigma_{{\bf k}}^{\mathrm{aux}}(\omega+i\eta)$ thus becomes
\begin{align}
\Sigma_{{\bf k}}^{\mathrm{aux}}(\omega+i\eta) & =\Sigma_{{\bf k}}^{\mathrm{ano}}(\omega+i\eta)+\frac{\Sigma_{{\bf k}}^{\mathrm{nor}}(\omega+i\eta)-\Sigma_{{\bf k}}^{\mathrm{nor}}(-\omega-i\eta)}{2}\nonumber\\
 & =\Sigma_{{\bf k}}^{\mathrm{ano}}(\omega+i\eta)+\frac{\Sigma_{{\bf k}}^{\mathrm{nor}}(\omega+i\eta)-\Sigma_{{\bf k}}^{\mathrm{nor}}(-\omega+i\eta)^{*}}{2}
\end{align}
with  
\begin{align}
\Sigma_{{\bf k}}^{\mathrm{nor}}(-\omega+i\eta)^{*} 
 & =s(\mathbf{k})+\sum_{\alpha}\frac{V_{\alpha}(\mathbf{k})^{2}\left[-\omega-i\eta+\epsilon_{f_{\alpha}}(\mathbf{k})\right]\left[\xi(\omega,{\bf k})-i2\omega\eta\right]}{\xi(\omega,{\bf k})^{2}+4\omega^{2}\eta^{2}}.
 \label{eq:sm_Snor_minus}
\end{align}
Now we plug Eqs.~(\ref{eq:sm_Snor}), (\ref{eq:sm_Sano}) and (\ref{eq:sm_Snor_minus}) 
into $\Sigma_{{\bf k}}^{\mathrm{aux}}(\omega+i\eta)$  
\begin{align}
\Sigma_{{\bf k}}^{\mathrm{aux}}(\omega+i\eta) & =D_{c}(\mathbf{k})+\sum_{\alpha}\frac{-V_{\alpha}(\mathbf{k})^{2}D_{f_{\alpha}}(\mathbf{k})\left[\xi(\omega,{\bf k})-i2\omega\eta\right]}{\xi(\omega,{\bf k})^{2}+4\omega^{2}\eta^{2}}\nonumber\\
 & +\left\{ \cancel{\frac{1}{2}s(\mathbf{k})}+\frac{1}{2}\sum_{\alpha}\frac{V_{\alpha}(\mathbf{k})^{2}\left[\omega+i\eta+\epsilon_{f_{\alpha}}(\mathbf{k})\right]\left[\xi(\omega,{\bf k})-i2\omega\eta\right]}{\xi(\omega,{\bf k})^{2}+4\omega^{2}\eta^{2}}\right\}\nonumber \\
 & -\left\{ \cancel{\frac{1}{2}s(\mathbf{k})}+\frac{1}{2}\sum_{\alpha}\frac{V_{\alpha}(\mathbf{k})^{2}\left[-\omega-i\eta+\epsilon_{f_{\alpha}}(\mathbf{k})\right]\left[\xi(\omega,{\bf k})-i2\omega\eta\right]}{\xi(\omega,{\bf k})^{2}+4\omega^{2}\eta^{2}}\right\}\nonumber \\
 & =D_{c}(\mathbf{k})+\sum_{\alpha}V_{\alpha}(\mathbf{k})^{2}\frac{\left[-D_{f_{\alpha}}(\mathbf{k})+\frac{\left[\omega+i\eta+\epsilon_{f_{\alpha}}(\mathbf{k})\right]-\left[-\omega-i\eta+\epsilon_{f_{\alpha}}(\mathbf{k})\right]}{2}\right]\left[\xi(\omega,{\bf k})-i2\omega\eta\right]}{\xi(\omega,{\bf k})^{2}+4\omega^{2}\eta^{2}}\nonumber\\
 & =D_{c}(\mathbf{k})+\sum_{\alpha}V_{\alpha}(\mathbf{k})^{2}\frac{\left[\omega-D_{f_{\alpha}}(\mathbf{k})+i\eta\right]\left[\xi(\omega,{\bf k})-i2\omega\eta\right]}{\xi(\omega,{\bf k})^{2}+4\omega^{2}\eta^{2}}\nonumber\\
 & =D_{c}(\mathbf{k})+\sum_{\alpha}V_{\alpha}(\mathbf{k})^{2}\frac{\left[\omega-D_{f_{\alpha}}(\mathbf{k})\right]\xi(\omega,{\bf k})+2\omega\eta^{2}}{\xi(\omega,{\bf k})^{2}+4\omega^{2}\eta^{2}}
  +i\eta\sum_{\alpha}V_{\alpha}(\mathbf{k})^{2}\frac{\left[\xi(\omega,{\bf k})-2\omega\left[\omega-D_{f_{\alpha}}(\mathbf{k})\right]\right]}{\xi(\omega,{\bf k})^{2}+4\omega^{2}\eta^{2}}.
\end{align}
The spectral weight of $\Sigma_{{\bf k}}^{\mathrm{aux}}(\omega+i\eta)$
is 
\begin{align}
A_{\Sigma}^{\mathrm{aux}}(\omega)=-\frac{1}{\pi}\mathrm{Im}\Sigma_{{\bf k}}^{\mathrm{aux}}(\omega+i\eta) & =-\frac{1}{\pi}\eta\sum_{\alpha}V_{\alpha}(\mathbf{k})^{2}\frac{\left[\xi(\omega,{\bf k})-2\omega\left[\omega-D_{f_{\alpha}}(\mathbf{k})\right]\right]}{\xi(\omega,{\bf k})^{2}+4\omega^{2}\eta^{2}}\nonumber\\
 & =-\frac{1}{\pi}\eta\sum_{\alpha}V_{\alpha}(\mathbf{k})^{2}\frac{\left[\omega^{2}-\eta^{2}-\epsilon_{f_{\alpha}}(\mathbf{k})^{2}-D_{f_{\alpha}}(\mathbf{k})^{2}-2\omega^{2}+2\omega D_{f_{\alpha}}(\mathbf{k})\right]}{\xi(\omega,{\bf k})^{2}+4\omega^{2}\eta^{2}}\nonumber\\
 & =-\frac{1}{\pi}\eta\sum_{\alpha}V_{\alpha}(\mathbf{k})^{2}\frac{-\left[\omega-D_{f_{\alpha}}({\bf k})\right]^{2}-\eta^{2}-\epsilon_{f_{\alpha}}(\mathbf{k})^{2}}{\xi(\omega,{\bf k})^{2}+4\omega^{2}\eta^{2}}\nonumber\\
 & =\frac{1}{\pi}\eta\sum_{\alpha}V_{\alpha}(\mathbf{k})^{2}\frac{\left[\omega-D_{f_{\alpha}}({\bf k})\right]^{2}+\eta^{2}+\epsilon_{f_{\alpha}}(\mathbf{k})^{2}}{\left[\omega^{2}-\eta^{2}-\epsilon_{f_{\alpha}}(\mathbf{k})^{2}-D_{f_{\alpha}}(\mathbf{k})^{2}\right]^{2}+4\omega^{2}\eta^{2}}>0,
\label{eq:Aw_aux}
\end{align}
and thus positive. When $\omega=\pm \sqrt{\epsilon_{f_{\alpha}}(\mathbf{k})^2+D_{f_{\alpha}}(\mathbf{k})^2}$, $A_{\Sigma}^{\mathrm{aux}}(\omega)$ diverges. Hence, $\pm \sqrt{\epsilon_{f_{\alpha}}(\mathbf{k})^2+D_{f_{\alpha}}(\mathbf{k})^2}$ are not only the poles of $\Sigma_{{\bf k}}^{\mathrm{nor}}$ and $\Sigma_{{\bf k}}^{\mathrm{ano}}$, but also those of $\Sigma_{{\bf k}}^{\mathrm{aux}}$.

\section*{SM2. Matsubara-frequency self-energy of K$_3$C$_{60}$ in the SC state}

For the SC state of K$_3$C$_{60}$ at $T=10$ K, Fig.~\ref{fig:Fig5} shows the self-energies $\Sigma^{\mathrm{nor}}(i\omega_n)$
and $\Sigma^{\mathrm{ano}}(i\omega_n)$ on the Matsubara frequency axis. The corresponding real-frequency counterparts are shown in 
Fig.~3 of the main text. 
The high-frequency tail exhibits considerable noise from the QMC sampling. 

\begin{figure*}[htp]
\includegraphics[clip,width=0.8\paperwidth,angle=0]{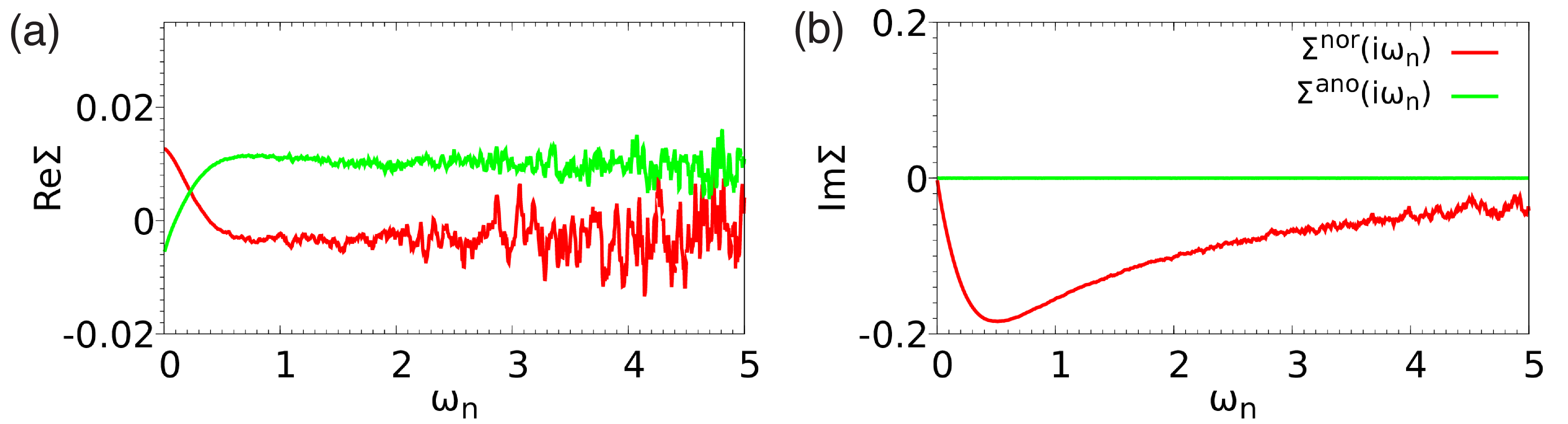}
\caption{
Matsubara-frequency self-energy of K$_3$C$_{60}$ at $T=10$ K in the SC state. The red (green) line shows
$\Sigma^{\mathrm{nor}}(i\omega_n)$ [$\Sigma^{\mathrm{ano}}(i\omega_n)$]. 
$\mathrm{Re}\Sigma^{\mathrm{nor}}(i\omega_n)$ (red line in panel (a)) is shifted downward by the Hartree-Fock value $\mathrm{Re}\Sigma^{\mathrm{nor}}_{\infty}$=1.82 for a better visualization. $\mathrm{Im}\Sigma^{\mathrm{ano}}(i\omega_n)$ is exactly zero (green line in panel (b)).
Here, we only show results for one of the degenerate $t_{1u}$ orbitals.
}
\label{fig:Fig5}
\end{figure*}

\section*{SM3.  Momentum-resolved spectral function $\log A({\bf k},\omega)$ and Hubbard bands}
 
Fig.~\ref{fig:Fig6} shows $\log A({\bf k},\omega)$ of K$_3$C$_{60}$ in a large energy window. The logarithmic scale is used to 
make the Hubbard bands more visible than it would be the case on a linear scale. 
 \begin{figure*}[htp]
\includegraphics[clip,width=0.8\paperwidth,angle=0]{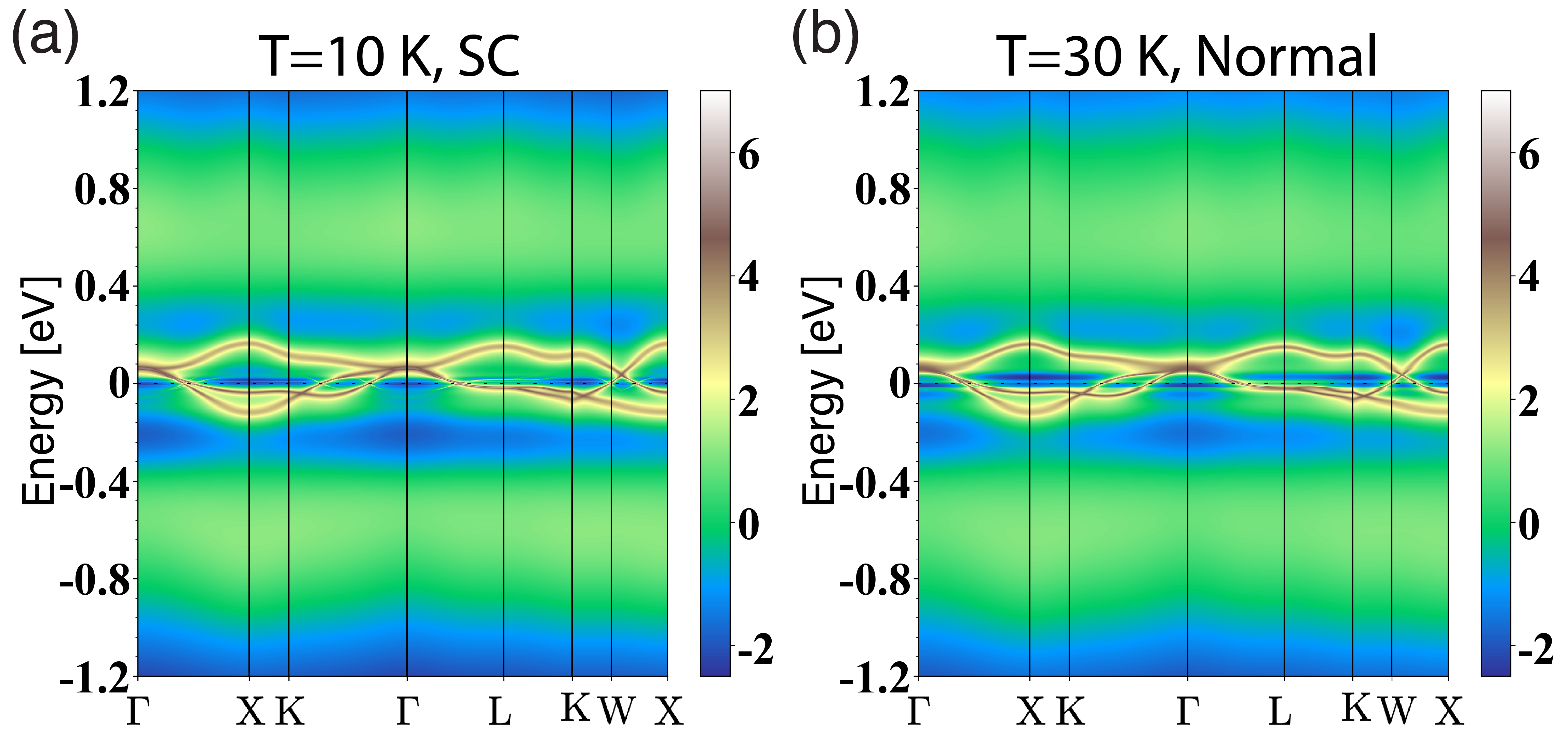}
\caption{
Momentum-resolved spectral function $\log A({\bf k},\omega)$ of K$_3$C$_{60}$ in a larger energy window.  
(a) $\log A({\bf k},\omega)$ for the superconducting state at $T=10$~K. 
(b) $\log A({\bf k},\omega)$ for the normal state at $T=30$~K.
}
\label{fig:Fig6}
\end{figure*}

\end{flushleft}
\end{widetext}
 
\end{document}